\documentclass[aps,pra,twocolumn,showpacs]{revtex4}
\input{epsf}
\usepackage{epsfig}

\begin{document}

\title{Transfer of trapped atoms between two optical tweezer potentials}

\author{Matthias Schulz$^{(\ddag)}$, Herbert Crepaz$^{(*)}$, Ferdinand Schmidt-Kaler$^{(\S)}$,
J{\"u}rgen Eschner$^{(*)}$, Rainer Blatt \vspace{3mm}}

\affiliation{ Institut f\"{u}r Experimentalphysik, Universit{\"a}t
Innsbruck, Austria \\
${(\ddag)}$ now at: Cassel Messtechnik GmbH, Dransfeld, Germany \\
${(\S)}$ now at: Institut f\"{u}r Quantenphysik, Universit{\"a}t
Ulm, Germany \\
${(*)}$ currently at: ICFO - Institut de Ciencies Fotoniques,
08860 Castelldefels (Barcelona), Spain}

\email{Juergen.Eschner@icfo.es}

\date{\today}

\begin{abstract}
Trapped, laser-cooled rubidium atoms are transferred between two
strongly focused, horizontal, orthogonally intersecting laser
beams. The transfer efficiency is studied as a function of the
vertical distance between the beam axes. Optimum transfer is found
when the distance equals the beam waist radius. Numerical
simulations reproduce well the experimental results.
\end{abstract}


\maketitle

\section{Introduction}

The spatial control of atoms, beyond their trapping in stationary
potentials, has been continuously gaining importance in
investigations of ultracold gases and in the application of atomic
ensembles and single atoms for cavity QED and quantum information
studies. Recent progress includes the trapping and control of
single atoms in dynamic potentials \cite{Grangier, Saffman}, the
sub-micron positioning of individual atoms with standing-wave
potentials \cite{Meschede, Chapman}, micro-structured and dynamic
traps for Bose-Einstein condensates \cite{Raizen, Foot} and, as
another example, the realisation of chaotic dynamics in
atom-optics "billiards" \cite{Milner2001, Davidson}.

The reaction of trapped atoms to dynamical variation of the
trapping potential is one central aspect of these developments.
Its understanding is important to design optimally the shape of
the potential and its temporal variations which provide the
desired control over the atoms and allow their manipulations. The
question of efficiently steering atoms by dynamically variable
light beams shares many similarities with the application of laser
beams for optical tweezers, for manipulating microscopic objects
such as beads or living cells \cite{Tweezers}. In the context of
quantum information processing, it is also related to recent
developments towards position control of single trapped ions in
complex, multi-segment ion traps~\cite{Ionshuttling}.

We report on a particular case of manipulation of atoms by dynamic
variation of light potentials formed by laser beams: the transfer
of a cold cloud of rubidium atoms between two horizontal,
intersecting, focused laser beams. The transfer happens by ramping
the intensities in the two beams, which is done slowly enough to
give all atoms time to adjust to the change. The measured signal
is the amount of atoms remaining in the second beam after the
first one has been fully switched off, i.e. the transfer
efficiency. Our main finding is a peculiar dependence of the
efficiency on the vertical distance between the two beams, which
we vary within about $\pm 3$ times the beam waist radius $w_0$.
The striking result is that we observe optimum transfer when the
beams are not fully overlapping but when their distance is about
$w_0$. For the situation with maximum overlap the transfer
efficiency is reduced by about a factor of two.

%Optimum transfer is observed when the distance is about $w_0$, but
%it is much less efficient when the beams overlap maximally.

The observed behaviour is reproduced in numerical simulations of
the situation, using classical trajectories of independent atoms.
Supported by these results, our explanation is that the finite
distance of $w_0$ between the beams is favourable for scattering
an atom from one beam to the other, because it does not create any
potential barrier, while at the same time the anharmonic potential
mixes the degrees of freedom of orthogonal directions.

\section{Experiment}

In the experiment, we trap a cloud of $^{87}$Rb atoms in a dipole
trap formed by about 0.7~W of light from a Ti:Sapphire laser at
810.0~nm, focused to a waist diameter of about $2w_0 = 15~\mu m$.
The atoms are transferred into the dipole trap from a standard
magneto-optical trap (MOT) formed by three retro-reflected beams
from a frequency-stabilised diode laser \cite{Schulz2002}. After
switching off the MOT, the atoms are held in the dipole trap for
100~ms, during which they thermalise. Then the second, orthogonal
laser beam is ramped up from 0 to to 0.4~W within 100~ms, followed
by ramping down the first beam to zero during the next 100~ms.
After another 220~ms of waiting time, the MOT beams are switched
back on, the dipole trap is switched off completely, and the
fluorescence of the recaptured atoms is measured during 100~ms.
The procedure is repeated several 10 times to average over
shot-to-shot fluctuations in the initial number of trapped atoms.

The two dipole trapping beams are horizontal and have the same
beam parameters. At their respective maximum powers of 0.7 and
0.4~W, the single-beam potentials are 150 and 86~MHz deep and have
42 and 32~kHz radial frequencies. Their ramping is done with
acousto-optical modulators (AOMs), while mechanical shutters are
used to switch them off completely before and after ramping. The
beams intersect at their focal points, apart from an adjustable
vertical distance which is the main experimental parameter.
Adjustment happens with a precision translation stage that moves
vertically the complete assembly of fiber collimator and focusing
lens which provides the second trapping beam.

\begin{center}
\begin{figure}[htb]
\epsfig{width=0.99\hsize, file=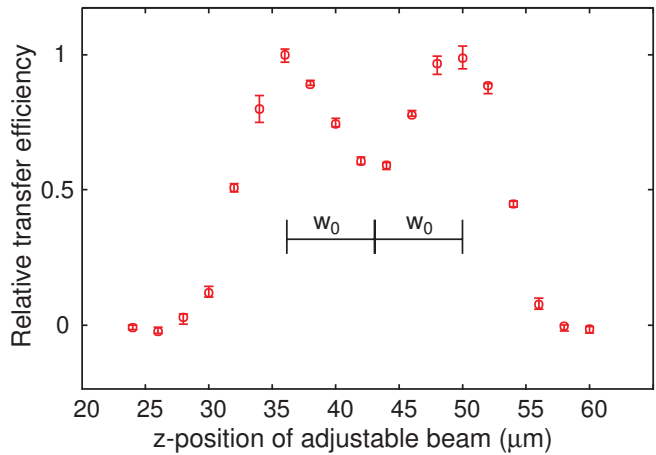}
\caption{Measured transfer efficiency between the two trapping
beams as a function of their vertical distance. The error bars
correspond to a confidence level of 68\%, determined from several
10 repetitions of the same measurement. The efficiency is
normalised to the maximum.} \label{Fig:ExpEfficiency}
\end{figure}
\end{center}

The main experimental result is shown in
Fig.~\ref{Fig:ExpEfficiency}. On variation of the vertical
distance, the transfer efficiency exhibits two maxima, around $\pm
w_0$. In the symmetric situation, at maximum overlap of the beams,
the efficiency is reduced, and, obviously, it goes to zero when
the beam distance becomes large.

\section{Qualitative description}

The probability of an atom to be transferred from one beam to the
other depends on the shape of the combined potential in the
crossing region. Here we provide a qualitative discussion of the
physical situation, before we present a numerical simulation of
the dynamics in the next section.

\begin{center}
\begin{figure}[htb]
\epsfig{width=0.99\hsize, file=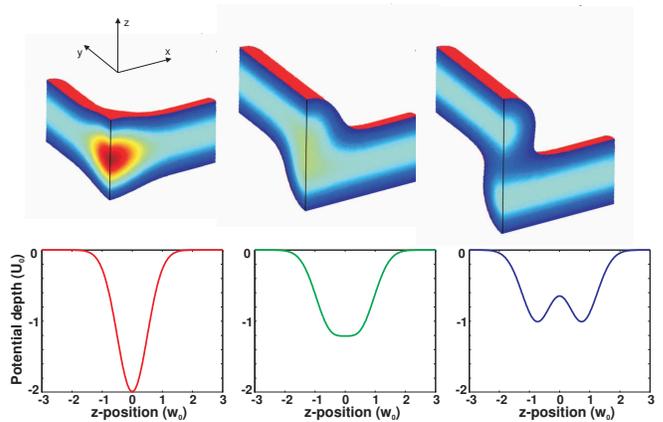}
\caption{Illustration of the potential in the crossing region, for
the vertical distance between the beams set to 0 (left), $w_0$
(center) and $1.5 w_0$ (right). Top: color- (or grey-scale-) coded
3-dimensional display. Bottom: vertical potential variation at the
center of the crossing region (along the black line in the top
display). } \label{Fig:Shapes}
\end{figure}
\end{center}

We denote the three spatial directions by $\hat{x}$ for the
propagation direction of the first beam, $\hat{y}$ for that of the
second, and $\hat{z}$ for the vertical direction. In
Fig.~\ref{Fig:Shapes}, we show three relevant cases of vertical
distance, assuming equal beam power. For zero distance the two
beams combine to form one localised 3-dimensional potential
"dimple" which is symmetric in $\hat{x}$ and $\hat{y}$, and
stronger in $\hat{z}$. For a distance larger than $w_0$, a
potential barrier separates the two beams. In the intermediate
case of a separation by $w_0$, neither a dimple nor a barrier is
formed. Instead, in the $\hat{z}$ direction the potential has a
non-Gaussian, flat-bottom shape.

Some qualitative conclusions can be drawn from these pictures. At
zero beam separation and for an atom with small kinetic energy,
the trajectory will not be affected significantly when it
traverses the beam crossing region, except for a faster radial
oscillation and some longitudinal acceleration and deceleration.
The potential is always close to harmonic and symmetric. Due to
this symmetry, the motion will remain centered around the axis of
the beam in which the atom enters the crossing region, and the
three directions of motion are not mixed. Thus a transfer between
the beams is not likely.

In the case of large beam separation, the potential barrier
impedes the transfer, so the probability that the atom changes
between the beams falls off to zero. The fall-off is expected to
happen faster (i.e.\ at smaller beam separation) for atoms with
lower energy.

The maximal transfer probability is observed in the intermediate
case. We ascribe this to the asymmetric potential in the crossing
region, which deflects all incoming atoms up- or downwards towards
the axis of the other beam, such that the motional degrees of
freedom become necessarily coupled. Moreover, the potential in the
crossing region is anharmonic, which enhances the mixing between
the directions.

\section{Model calculations}

Numerical simulations of the classical motion of a particle in the
crossed-beam potential serve as a complementary approach to
understand the observed dynamics. Based on the analytical
expressions for the potential and its spatial gradient, the
differential equations of motion were solved numerically using
standard tools \cite{Matlab}. At typical atom temperatures of
50~$\mu$K and densities of several 10$^{10}$~cm$^{-3}$, the
classical trajectory approach is well justified. Instead of
simulating the complete transfer process, which would have
required significantly more computational resources, we looked at
the dynamics of a single particle in the potential formed by two
beams of constant, equal power. Hence these calculations highlight
the effect of the coupling between the motional degrees of freedom
in the crossing region, and of the potential barrier (when it
exists) between the beams.

\begin{center}
\begin{figure}[htb]
\epsfig{width=0.99\hsize, file=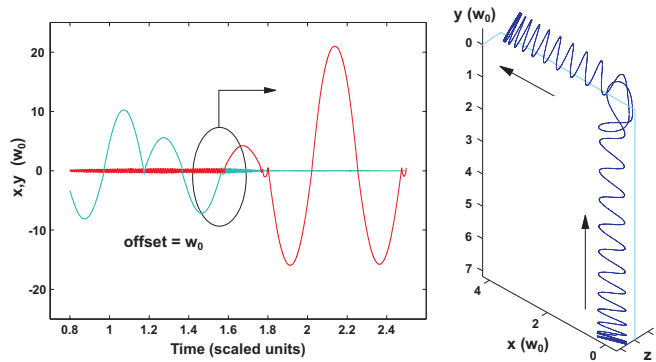} %
\caption{Sample trajectory from the numerical simulation. Left:
excursion in $\hat{x}$ and $\hat{y}$, in units of $w_0$, vs.\
time; $\hat{z}$-motion is not shown. Right: 3-dimensional display
zooming into the circled region of the left plot which shows a
transfer event. } \label{Fig:Trajectory}
\end{figure}
\end{center}

Figure \ref{Fig:Trajectory} shows a sample trajectory, for a
displacement between the beams equal to the beam waist $w_0$. The
numerical calculations were programmed with high temporal
resolution to account for the strongly different trap frequencies
and for the anharmonic potential encountered by particles off the
beam axes. In the simulations the total energy (the sum of kinetic
and potential energy) was found to change slowly with time, due to
accumulated numerical errors, but it was practically constant
during a single transit through the crossing region, such that it
served as a parameter characterising each transit event. The
probability for a transfer from one beam to the other during a
transit through the crossing region was then recorded as a
function of the energy and the beam separation. A particle was
considered to be transferred between the beams if it entered the
central region from a distance larger than $3w_0$ in one beam, and
left this region in the other beam to a distance larger than
$3w_0$. A transit was defined as any event where the particle
entered and left the crossing region in either of the beams. The
transfer probability is the ratio of the number of transfers to
the number of transits. The initial conditions were chosen at
random; gravity was neglected \cite{FootnoteGravity}.

\begin{center}
\begin{figure}[htb]
\epsfig{width=0.99\hsize, file=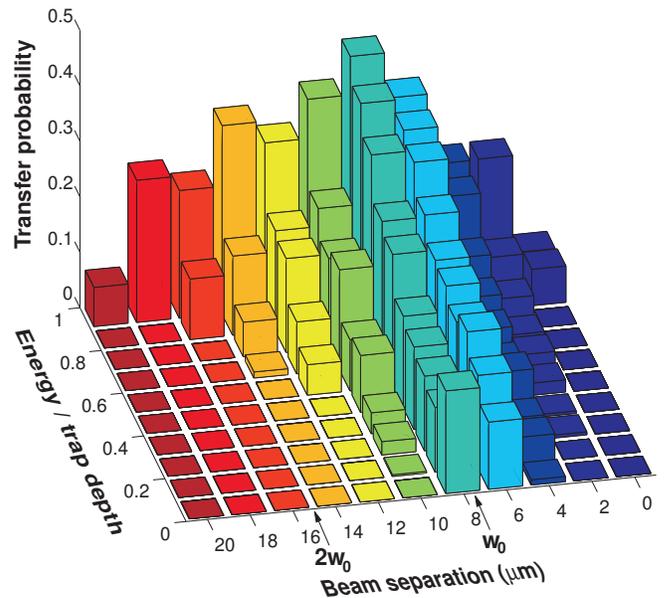} %
\caption{Simulated transfer efficiency vs.\ beam separation, at
different total energy of the particle. } \label{Fig:EffEnergy}
\end{figure}
\end{center}

Figure \ref{Fig:EffEnergy} shows a histogram of transfer
probability vs. beam separation and particle energy from such a
numerical investigation. The value of each bin is based on 100 to
1000 transit events.

Its general features agree well with the qualitative explanation
given earlier. At small beam separation, significant transfer only
happens at high particle energies. At beam separations around
$w_0$, the transfer probability is more equally spread over all
possible energies, and generally much higher than in the case of
small beam separation. Beam separations exceeding $w_0$ show no
transfer at low particle energies, due to the potential barrier
between the beams; the excluded energy range grows with increasing
separation.

\begin{center}
\begin{figure}[htb]
\epsfig{width=0.99\hsize, file=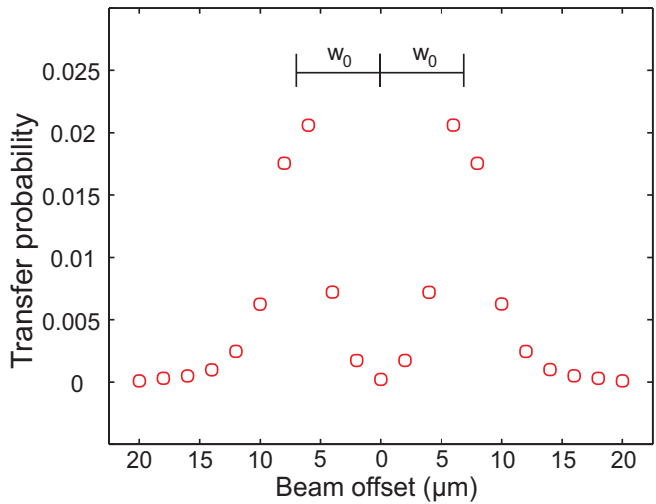} %
\caption{Simulated transfer efficiency vs.\ beam separation, for a
thermal distribution of particle energies as in the experiment.
This diagram must be compared to Fig.~\ref{Fig:ExpEfficiency}. }
\label{Fig:SimEfficiency}
\end{figure}
\end{center}

Finally, Fig.~\ref{Fig:SimEfficiency} shows the simulated transfer
probability vs.\ beam separation for a thermal distribution of
atoms; this diagram must be compared with the experimental result
of Fig.~\ref{Fig:ExpEfficiency}. It was calculated from the data
of Fig.~\ref{Fig:EffEnergy} by weighting them with a thermal
distribution of particle energy in a 3-dimensional harmonic
oscillator. The temperature was taken to be 10\% of the trap
depth, which we determined in independent measurements, and which
has also been consistently found in several other experiments
\cite{TrapEnergy}.

The simulated transfer probability is maximal around a beam
separation of $w_0$, with a pronounced minimum at the position of
perfect crossing of the beams. For separations larger than $w_0$,
the probability falls off steeply. Therefore, the main features of
the measurement are already well reflected in these simple
simulations, using a static trap potential of Gaussian beams. The
largest deviation between experiment and simulation is observed
for small beam separation, where the measured transfer probability
is significantly higher. We attribute this to the transient
situation of non-equal intensities of the beams during ramp-up and
ramp-down in the experiment, which is less symmetric and may
therefore mix the motion more efficiently. Other effects, such as
elastic collisions between atoms in the crossing region, may also
play a role in redistributing energy among the three modes of
oscillation, and thus enhance the transfer efficiency at small
beam separation for a cloud of atom, as in the experiment,
relative to the simulated single-particle case. We exclude, on the
other hand, that experimental inaccuracies, such as non-Gaussian
beam shapes or beam pointing instabilities, have broadened the
central part of the measured efficiency curve, because no such
broadening is observed on the outer wings of the curve.

\section{Conclusions}

In summary, we have measured the transfer of trapped atoms between
two crossed laser beams, varying the distance between the beams.
We find optimum transfer at non-zero beam separation, when the
mixing between the motional degrees of freedom is favoured by the
anharmonic potential, while no potential barrier is formed between
the beams. The main characteristics of the experimental
observations are reproduced in numerical simulations of the
dynamics. They become visible already in the situation of two
crossed beams of constant, equal power. Our results are relevant
for designing efficient loading mechanisms for optical traps and
optical lattices, and they may find applications in optical
tweezer technology for nano- and micro-particles.

\acknowledgements{This work was supported in parts by the Austrian
Science Fund (FWF, Project SFB15) and by the European Commission
(QGATES, IST-2001-38875).}


\begin{thebibliography}{}
\bibitem{FootnoteGravity}
The optical potentials within the focus region of about 10~$\mu$m
size are at a level of tens of MHz, such that gravity
($\sim$2.5~kHz/$\mu$m) can be ignored for a principal
understanding of the dynamics.
\end{thebibliography}
\end{document}